\documentclass{iopart}

\usepackage{url}
\usepackage{amsbsy}
\usepackage{iopams}
\usepackage{graphicx}

\newcommand{\eqref}[1]{{(\ref{#1})}}
 
\begin{document}

\title{Sensitivity and parameter-estimation precision for alternate LISA configurations}

\author{Michele Vallisneri, Jeff Crowder and Massimo Tinto}
\address{Jet Propulsion Laboratory, California Institute of Technology, Pasadena, CA 91109}

\ead{Michele.Vallisneri@jpl.nasa.gov}

\begin{abstract}
  We describe a simple framework to assess the LISA scientific
  performance (more specifically, its sensitivity and expected
  parameter-estimation precision for prescribed gravitational-wave
  signals) under the assumption of failure of one or two
  inter-spacecraft laser measurements (\emph{links}) and of one to
  four intra-spacecraft laser measurements.
  We apply the framework to
  the simple case of measuring the LISA sensitivity to monochromatic
  circular binaries, and the LISA parameter-estimation precision for
  the gravitational-wave polarization angle of these systems.
  Compared to the six-link baseline configuration, the five-link case
  is characterized by a small loss in signal-to-noise ratio (SNR) in the
  high-frequency section of the LISA band; the four-link case shows a
  reduction by a factor of $\sqrt{2}$ at low frequencies, and by up to
  $\sim 2$ at high frequencies. The uncertainty in the estimate of
  polarization, as computed in the Fisher-matrix formalism, also
  worsens (but only for sources above 1 mHz) when moving from six to five,
  and then to four links: this can be explained entirely by 
  the reduced SNR available in those configurations.
  In addition, we prove (for generic signals) that the SNR and Fisher matrix
  are invariant with respect to the choice of a basis of TDI observables;
  rather, they depend only on which inter-spacecraft and intra-spacecraft
  measurements are available.
\end{abstract}

\vspace{-12pt}


\section{Introduction}

LISA (the Laser Interferometer Space Antenna) is a deep-space mission planned jointly by the National Aeronautics and Space Administration and the European Space Agency. LISA seeks to detect and study gravitational waves (GWs) in the mHz frequency band. It consists of three spacecraft flying in a triangular formation, whose relative positions will be monitored by way of laser interferometry \cite{PPA98}.
In contrast to ground-based interferometric GW detectors, LISA will have multiple readouts,
for the six laser links between the spacecraft (one in each direction across each arm).
These data streams, properly time shifted and linearly combined, provide observables that are insensitive to laser-frequency fluctuations and optical-bench motions, and that have different couplings to GWs and to the remaining system noises \cite{TD05}. This technique is known as Time-Delay Interferometry
(TDI). Early on it was realized that different TDI observables can be built from different subsets of the six inter-spacecraft laser measurements \cite{ETA00}, thus providing failure resistance against the loss of one or more measurements; furthermore, two or three linearly independent TDI observables can be used together to increase SNR and improve GW-parameter determination (this insight goes back at least to \cite{Cutler98}).

In this paper we pull together and complete the results scattered throughout the TDI literature (see references in \cite{TD05}) to build a simple framework that can assess the multi-observable LISA science performance when all measurements are available, and in failure scenarios where up to two of the six LISA inter-spacecraft measurements and up to four of the six intra-spacecraft measurements are lost. (We do not analyse the case of
\emph{three} lost inter-spacecraft laser-links: in this scenario, LISA
cannot make wide-band measurements of gravitational radiation, but
narrow-band measurements are still possible at frequencies equal to
integer multiples of the inverse round-trip light time between the two
spacecraft that preserve bidirectional laser measurements
\cite{Tinto98}.)

In section \ref{sec:concept} we introduce our notation and our model of the LISA measurement; in \ref{sec:observables}, we describe a simple linear-algebra procedure to obtain the TDI observables that can be constructed in each LISA configuration; in \ref{sec:invariance}, we prove that any choice of a basis of observables can be used to compute the SNR and Fisher matrix, which depend only on the available measurements; in \ref{sec:bases}, we give explicit expressions for the \emph{noise-orthonormal} bases that simplify these computations; in \ref{sec:sensitivity} and \ref{sec:precision}, we apply our framework to the problem of determining the six-link, five-link, and four-link LISA sensitivity to sinusoidal signals and its precision (i.e., the expected statistical error due to instrument noise) in the determination of the polarization of monochromatic binaries, thus quantifying the heuristic statement that additional TDI observables help disentangle the GW polarization states.

\section{The LISA noise response}
\label{sec:concept}

We adopt the ``classic'' conceptualisation of the LISA measurement
developed by Armstrong, Estabrook, and Tinto (see especially
\cite{ETA00}), whereby: (i) each LISA spacecraft contains two
optical benches, each with a proof mass and a laser; (ii) the
inter-spacecraft one-way Doppler measurement ``$y$'' on each bench consists of the
fractional frequency difference between the incoming laser (bounced
off the local proof mass) and the local laser (unbounced); (iii) the
intra-spacecraft measurement ``$z$'' on the same bench consists of the
fractional frequency difference between the local laser (unbounced)
and the laser from the other bench on the spacecraft, transmitted via
optical fibre (and bounced on the other bench's proof mass). The noise
in each of the measurements is modelled as due to the frequency
fluctuations of the six lasers, to the random displacements of the six
proof masses and of the six optical benches, and (for the $y$'s) to
shot noise and other optical-path noise in the low-SNR
inter-spacecraft laser links. The optical-fibre noise can be removed
by using the $z$'s only in differences of the $z$'s on the same bench;
and the optical-bench motions along the lasers' lines of sight can be
absorbed into the laser frequency noises. Without loss of generality,
we assume also that the central frequencies of the lasers are all the
same, and that all the Doppler beat notes due to the relative motions
of the spacecraft have been removed via heterodyne measurements
\cite{TEA02, TAE07}.

\begin{figure}
\includegraphics[width=0.97\textwidth]{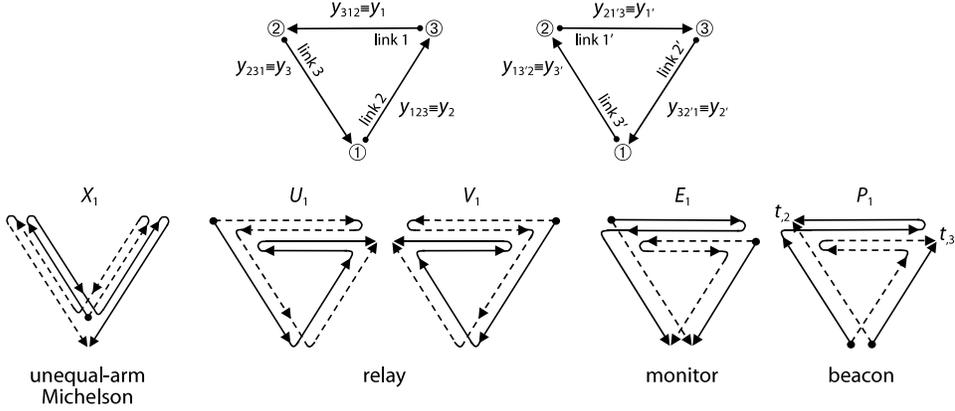}
\caption{Top panel: the LISA inter-spacecraft measurements. 
  Each arrow represents the frequency-comparison measurement $y_l
  \equiv y_{slr}$ taken on the spacecraft at the tip of the arrow
  between the incoming laser (moving in the direction of the arrow and
  experiencing the light time $L_l$) and the local laser. Unprimed
  (primed) link indices correspond to counter-clockwise (clockwise)
  laser propagation.  Bottom panel: first-generation TDI observables
  as synthesized interferometers \cite{geotdi}. The four kinds of
  four-link observables are distinguished by the direction of the
  links converging on one spacecraft (in this case 1), which makes it
  a ``centre,'' ``relay,'' ``monitor,'' or ``beacon.''
\label{fig:geometry}}
\vspace{-6pt}
\end{figure}

Using the notation of \cite{synthlisa} and \cite{geotdi}, the $y$ and $z$ noise responses
are given by
\begin{eqnarray}
\fl y^\mathrm{noise}_{slr}(t) = 
\left\{
\begin{array}{l}
C^*_s\bigr(t - L_l(t)\bigl) - C_r(t) + y^\mathrm{op}_{slr}(t) - 2 pm_r(t)
\quad \mbox{for unprimed $l$}, \\[3pt]
C_s\bigr(t - L_l(t)\bigl) - C^*_r(t) + y^\mathrm{op}_{slr}(t) - 2 pm^*_r(t)
\quad \mbox{for primed $l$},
\end{array}
\right.
\label{eq:ynoises}
\end{eqnarray}
and
\begin{eqnarray}
\fl z^\mathrm{noise}_{slr}(t) = 
\left\{
\begin{array}{l}
C^*_r(t) - C_r(t) + 2 pm^*_r(t)
\quad \mbox{for unprimed $l$}, \\[3pt]
C_r(t) - C^*_r(t) + 2 pm_r(t)
\quad \mbox{for primed $l$},
\end{array}
\right.
\label{eq:znoises}
\end{eqnarray}
where $y_{slr}$ denotes the inter-spacecraft measurement obtained for
the laser incoming from spacecraft $s$ into spacecraft $r$, and
travelling along link $l$ (see figure \ref{fig:geometry}; each link $l$ sits
across from spacecraft $l$ in the LISA triangle, and takes a prime if $slr$ is an odd permutation of $123$); furthermore, $z_{slr}$ denotes the intra-spacecraft
measurement on the same bench as $y_{slr}$ (this traditional notation is somewhat
unfortunate, since the indices $s$ and $l$ refer to distant spacecraft that
have no role in the local measurement).
In these equations, the $C_s$ and $C^*_s$ represent the six laser frequency noises, the $pm_r$ and
$pm^*_r$ the six proof-mass velocity fluctuations along the lasers' lines of
sight, the $y^\mathrm{op}_{slr}$ the six inter-spacecraft optical-path
noises, and $L_l$ is the one-way light time along link $l$ (the
asterisks over the $C$ and $pm$ \emph{do not} denote complex
conjugation, but rather different noise variables).

TDI observables are linear combinations of several $y$ and $z$,
appropriately time delayed so that all instances of laser frequency noise ($C_r$ and $C^*_r$)
cancel out.  It is useful to introduce time-delay operators
$\mathcal{D}_l$ such that for any measurement $x(t)$,
\begin{equation}
\fl \mathcal{D}_l x(t) \equiv x\big(t - L_l(t)\big), \quad
\mathcal{D}_m \mathcal{D}_l x(t) \equiv x\Big(t - L_l(t) - L_m\big(t - L_l(t)\big)\Big), \quad \ldots
\end{equation}
and so on, with the shorthand $x_{;l}(t) \equiv \mathcal{D}_l x(t)$,
$x_{;ml}(t) \equiv \mathcal{D}_m \mathcal{D}_l x(t)$, and so on. The
time-delay operators commute only if the $L_l$ are not functions of
time, in which case the delays are usually denoted by subscripts set off by commas.
For simplicity, in this paper we shall concern ourselves only with this
case, which corresponds to ``modified'' TDI observables.
(``Second-generation'' TDI observables are necessary to remove laser
noise completely because of the ``breathing'' of the LISA arms \cite{cornhell,TEA04};
however, when compared to the corresponding first-generation
TDI observables, they engender only negligible corrections to the GW
and secondary-noise responses, which are the building blocks of the
SNR and Fisher matrix. Moreover, the second-generation observables can
be approximated as finite-difference derivatives of their
first-generation versions \cite{TEA04,geotdi}, and this effect factors
out in the computation of the SNR and Fisher matrix. For these
reasons, the results of this paper remain valid for second-generation
TDI.)

The problem of finding combinations of the $y$'s and $z$'s that cancel
the six laser noises, $C_s$ and $C^*_s$, can be reduced \cite{TEA04} to
the problem of finding combinations of the new
variables\footnote{These substitutions are essentially equivalent, yet
  slightly different from those introduced in \cite{TEA04}, with the
  advantage of leading to more symmetric noise responses.}
\begin{equation}
y'_{slr} = y_{slr} + {\textstyle \frac{1}{2}}(z_{rl's,l} - z_{slr}),
\label{eq:simplifiedy}
\end{equation}
which cancel the \emph{three} equivalent laser noises $C'_s \equiv
(C_s + C^*_s)/2$. This reduction removes the $z$'s from
consideration,\footnote{It would seem that working with the $y'$ it
  becomes hard to enforce the constraint that the $z$ be used only in
  the same-spacecraft differences $z_{231} - z_{32'1}$, $z_{312} - z_{13'2}$, $z_{123}
  - z_{21'3}$ (on spacecraft 1, 2 and 3 respectively), but this happens
  naturally in TDI combinations because of the way that the laser
  noises enter the $y'$ noise responses. This is easy to see from the
  viewpoint of geometric TDI \cite{geotdi}, where $y'$ arrows that
  begin or end at the same spacecraft always appear in the sequences
  \includegraphics[height=10pt]{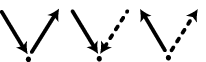} (which involve the
  $z$ difference at the spacecraft) or
  \includegraphics[height=10pt]{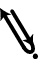} (which involves
  neither of the $z$'s at the spacecraft).\label{note:geotdi}} and
leads to the noise responses
\begin{eqnarray}
\fl {y'}^{\mathrm{noise}}_{slr}(t) = 
\left\{
\begin{array}{l}
C'_{s,l} - C'_r + y^\mathrm{op}_{slr} - 2 pm_r + pm_{s,l} - pm^*_r
\quad \mbox{for unprimed $l$}, \\[3pt]
C'_{s,l} - C'_r + y^\mathrm{op}_{slr} - 2 pm^*_r + pm^*_{s,l} - pm_r
\quad \mbox{for primed $l$}.
\end{array}
\right.
\label{eq:newynoises}
\end{eqnarray}
For terseness, in the rest of this paper we shall contract each of the index triples
$312$, $123$, $231$, $21'3$, $32'1$, and $13'2$ to its middle link
index alone, as shown at the top of figure \ref{fig:geometry}, so that
$y_l \equiv y_{slr}$ and $z_l \equiv z_{slr}$.

The recently proposed \cite{bonny} LISA architecture with a single proof
mass on each spacecraft and with ``strap-down'' measurements (whereby
the lasers are not bounced on the proof masses, but additional
measurements are taken of the position of the latter with respect to
the optical benches) would lead to analogous, but different noise
responses. We expect the results of this paper to persist in the new
architectures with slight modifications of \ref{eq:simplifiedy}, but
careful verification is certainly indicated once the LISA architecture
is finalized.

\section{TDI observables from linear algebra and the LISA failure modes}\label{TDI_algebra}
\label{sec:observables}

When the delay operators commute, they have an especially useful
representation in the Fourier domain,
\begin{equation}
\mathcal{D}_l \tilde{x}(f) = \Delta_l \tilde{x}(f), \quad
\mathcal{D}_m \mathcal{D}_l \tilde{x}(f) = \Delta_m \Delta_l \tilde{x}(f), \quad \ldots
\end{equation}
where $\Delta_l = \exp^{2 \pi \rmi f L_l}$. This representation turns
the search for laser-noise--cancelling combinations into a problem of
linear algebra. The laser-noise content of the $y'_l$ measurements is
described by the equation
\begin{equation}
\label{eq:ytoc}
\tilde{\mathbf{y}}' = D_6 \tilde{\mathbf{C}}',
\end{equation}
where $\tilde{\mathbf{y}}'$ is the vector $[y'_1,y'_2,y'_3,y'_{1'},y'_{2'},y'_{3'}]^T$,
$\tilde{\mathbf{C}}' \equiv [C'_1,C'_2,C'_3]^T$, and
\begin{equation}
D_6 = \left(\!\!
\begin{array}{ccc}
0 & -1 & \Delta_1 \\
\Delta_2 & 0 & -1 \\
-1 & \Delta_3 & 0 \\
0 & \Delta_{1'} & -1 \\
-1 & 0 & \Delta_{2'} \\
\Delta_{3'} & -1 & 0
\end{array}\!\!\right);
\end{equation}
now, a generic observable (not necessarily laser-noise--cancelling) is
a linear combination $\mathbf{a}^T \tilde{\mathbf{y}}'$ (where
``${}^T$'' denotes vector transposition), where the coefficients
$\mathbf{a}$ will normally be polynomials of the $\Delta_l$ with
integer coefficients. It follows from \eqref{eq:ytoc} that the
$\mathbf{a}$ that correspond to laser-noise--cancelling observables
must span the null space of $D_{6}^T$,
\begin{equation}
\mathbf{a}^T \! D_6 \tilde{\mathbf{C}}' = 0 
\quad \Rightarrow \quad D_{6}^T \mathbf{a} = 0;
\end{equation}
thus, the task of finding TDI observables reduces to finding a basis
for $\mathrm{null}(D_6^T)$, and \emph{all} TDI observables can then be
represented as linear combinations of the basis elements (although
such representations may involve $\Delta_l$'s, and therefore delays,
as we shall see in section \ref{sec:fivelink}). The Fourier-domain
representation is especially suited to our goals of assessing the
reduction in SNR associated with particular failure
configurations, and of evaluating the parameter-estimation precision
for observed gravitational signals, since these operations involve
calculations performed in the Fourier domain.

The loss of one or more inter-spacecraft measurements results in the
loss of the corresponding $y'_l$. However, the linear-algebra problem
of \eqref{eq:ytoc} can still be formulated after removing the
appropriate rows in $\tilde{\mathbf{y}}'$ and $D_{6}$. In general, if
$n$ is the number of measurements and $m$ the number of laser noises
to be cancelled, and if the subsetted $D_{6}$ has full rank (which is
generally the case), it follows that the dimension of
$\mathrm{null}(D^T)$ is $r = n - m$.  Thus, for ``full LISA,'' with
six inter-spacecraft measurements and three noises, we can construct
three independent TDI observables\footnote{Although a combination of
  the three is insensitive to GWs in the long-wavelength limit because
  of symmetry; see figure \ref{fig:sensitivity}.}; these are reduced
to two after one measurement is lost, and to one after two are lost.
In section \ref{sec:bases} we describe which of the standard TDI observables can be reconstructed in each case.

The effects of losing one or more intra-spacecraft measurements are
subtler. Consider for instance losing $z_{231} \equiv z_3$ (on spacecraft 1): from \eqref{eq:simplifiedy}, we see that we immediately lose $y'_3$ and $y'_{3'}$; we also lose the ability of using $y'_2$ and $y'_{2'}$, which involve the other measurement on spacecraft 1, $z_{32'1} \equiv z_{2'}$ (remember that $z_3$ and
$z_{2'}$ can appear only in the combination $z_3 - z_{2'}$ if fibre
noise must be cancelled; thus, losing one or two $z$ measurements on
the same spacecraft is equivalent).  However, it turns out that the
combinations $y'_{3'} + y'_{3,3'}$ and $y'_{2} + y'_{2',2}$ (in the
graphic notation of geometric TDI \cite{geotdi},
\includegraphics[height=12pt]{geotdi-paths-b-2} and
\includegraphics[height=12pt]{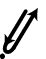}) contain neither $z_3$
nor $z_{2'}$ (and also no $C'_1$!). Thus we have four available
measurements and two noises to be cancelled, so two independent TDI
observables can be constructed. If some or all $z$ measurements are
lost on two spacecraft (say, 2 and 3) then only two combined
measurements are available (in this case, $y'_{3} + y'_{3',3}$ and
$y'_{2'} + y'_{2,2'}$) with one laser noise to cancel, so a single
independent TDI observable can be constructed (it is the unequal-arm
Michelson centred on the non-faulty spacecraft). If both inter- and
intra-spacecraft measurements are lost, then at most one TDI
observable can be constructed (and it is always an unequal-arm
Michelson), and in some cases none. Again, see section \ref{sec:bases}. It is also possible to come to the same
conclusions by working with a larger linear system that involves the
available $y_l$ (not $y'_l$), $z_l$ and the six laser noises $C_r$ and
$C^*_r$.

\section{Invariance of SNR and Fisher matrix by change of TDI basis}
\label{sec:invariance}

Of course, the basis of $\mathrm{null}(D^T)$ (with $D$ the appropriate
row-reduced matrix for the available measurements) is not unique;
however, we now show that the total SNR and Fisher matrix are
invariant with respect to the choice of a TDI basis, so \emph{they
  depend only on which $y_l$ and $z_l$ measurements are available},
and can computed over any convenient basis. From the SVD decomposition
\cite{golub}
\begin{equation}
\underbrace{D^T}_{m \times n} = \underbrace{U}_{m \times m} \cdot \underbrace{\Sigma}_{m \times n} \cdot \underbrace{V^\dagger}_{n \times n} \quad \mbox{(with $\Sigma_{ij} = 0$ for $i \neq j$)},
\end{equation}
it follows that a basis for $\mathrm{null}(D^T)$ is given by the
columns of $V$ that correspond to the ``extra'' $r$ null columns of
$\Sigma$.  The SVD decomposition is degenerate with respect to
nonsingular linear combinations of these columns (i.e., with respect
to choosing a different TDI basis).  Calling $W$ the $n \times r$
matrix of these columns, the optimal matched-filtering SNR for a
source $\mathbf{h}(f)$ whose components are given by the individual
TDI observables $X_\alpha = y'_i(t) W_{i\alpha}$ is given by
\begin{eqnarray}
\mathrm{SNR}^2 & = 4 \, \mathrm{Re} \int_0^{+\infty}
h_\alpha^\dagger(f) S^{-1}_{\alpha \beta}(f) h_\beta(f) \, \rmd f \\
& = 4 \, \mathrm{Re} \int_0^{+\infty} \mathbf{h}^\dagger(f) [W^T C(f) W^*]^{-1} \mathbf{h}(f) \, \rmd f, \nonumber
\end{eqnarray}
since
\begin{eqnarray}
\fl \frac{1}{2} S_{\alpha\beta}(f) \delta(f - f') & = \langle [y'_i(f) W_{i\alpha}] [y'_j(f') W_{j\beta}]^* \rangle =
W_{i\alpha} \langle y'_i(f) {y'}^*_j(f') \rangle W^*_{j\beta} \\ \fl & =
\frac{1}{2} W^T C(f) W^* \delta(f - f'), \nonumber
\end{eqnarray}
where the $C_{ij}$ components are the secondary-noise cross-spectra of the individual $y'_i$.
As discussed in \ref{app:snr}, $C_{ij}$ matrix is in general complex and Hermitian. 

If we change the TDI basis by means of a nonsingular linear
transformation $\mathbf{h} \rightarrow B \mathbf{h}$ (so that we are
computing $S_{\alpha' \beta'}$ over the TDI observables $X_\alpha' =
B_{\alpha' \alpha} X_\alpha = B_{\alpha' \alpha} y'_i(t) W_{i\alpha}$)
we get
\begin{eqnarray}
\fl
\mathbf{h}^\dagger [W^T C W^*]^{-1} \mathbf{h} \rightarrow
\mathbf{h}^\dagger B^\dagger [B W^T C W^* B^\dagger]^{-1} B \mathbf{h} & =
\mathbf{h}^\dagger B^\dagger (B^\dagger)^{-1} [W^T C W^*]^{-1} B^{-1} B \mathbf{h} \nonumber \\ & =
\mathbf{h}^\dagger [W^T C W^*]^{-1} \mathbf{h},
\label{eq:prodsn}
\end{eqnarray}
so the SNR is invariant. Indeed, any noise inner product
$(\mathbf{h},\mathbf{g})$, and therefore the Fisher matrix elements
$(\mathbf{h}_{,\mu},\mathbf{h}_{,\nu})$, are also invariant under
similar transformations.

As a corollary we get the intuitive fact that computing the SNR
with an extra TDI observable that is a linear combination of the $r$
others (which amounts to a $(r+1) \times r$ matrix $B$) does not
change the result.  To see this, we inject an additional $(r+1) \times
(r+1)$ linear transformation $B'$ that makes one of the $(r+1)$ TDI
observables identically null; the entire product \eqref{eq:prodsn} can
then be rewritten by dropping one of the rows of $B'B$, so we are back
to the case of a nonsingular $r \times r$ transformation.

The equivalence of different TDI bases for the purpose of computing the Fisher matrix can also be proved with a different approach, closer to the reasoning of Romano and Woan \cite{tdiconnection}: we consider the estimation problem of determining the GW parameters \emph{and} the laser frequency noises in the presence of secondary instrument noise, assumed Gaussian and stationary. The likelihood of the LISA data is then written in terms of the $y'_i$ and their secondary-noise cross-spectra, and (without loss of generality) rewritten in a TDI/non-TDI basis where the first $m$ components span $\mathrm{null}(D^T)$ and the other $r$ span $\mathrm{null}(D^T)^\bot$. We compute the Fisher matrix by taking derivatives of the log likelihood with respect to the GW parameters and to the individual frequency components of laser noise (again, our premise is that we are estimating these in addition to the GW parameters, even if in practice it will turn out that only some combinations of the laser noises can truly be estimated from the LISA measurements).
Using the Frobenius--Schur formula \cite{bodewig}, we can then show that the GW-parameter sector of the inverse Fisher matrix (i.e., the expected covariance matrix for the GW-parameter estimators) is equal to the inverse of the TDI sector of the full Fisher matrix, with all the laser-noise components dropping out. Since the projection on the $\mathrm{null}(D^T)$ and $\mathrm{null}(D^T)^\bot$ subspaces can be cast as a geometrical operation, we see that the Fisher matrix does not depend on the choice of TDI observables, but only on the geometry of $\mathrm{null}(D^T)$.

\section{Noise-orthogonal bases of TDI observables}
\label{sec:bases}

An especially useful linear transformation of a TDI basis is the one
that leads to an orthonormal basis of $S_{\alpha \beta}$, since the
SNR then has the simplified form
\begin{equation}
\label{eq:simplesnr}
\mathrm{SNR}^2 = \sum_{\alpha'} 4 \, \mathrm{Re} \int_0^{+\infty}
\frac{{h_{\alpha'}}^\dagger(f) h_{\alpha'}(f)}{S_{\alpha'\alpha'}(f)} \rmd f;
\end{equation}
a similar expression follows for the Fisher-matrix elements. We now
work out such bases and the corresponding $S_{\alpha'\beta'}$ under
various LISA failure modes.
\subsection{Four-link configurations}

We begin with the case of two lost $y_l$ measurements, where
$\mathrm{null}(D^T)$ has dimension one, so that essentially only one
TDI observable can be constructed, which is trivially a
noise-orthonormal basis.  Depending on which measurements are lost,
this observable is one of the standard four-link observables
(unequal-arm--Michelson, relay, beacon, and monitor) described in
\cite{TEA04}, and shown pictorially in the bottom part of figure
\ref{fig:geometry}.  In this section we spell them out and identify
them with a new compact naming scheme.

If the missing $y_l$ are along the same arm (say, $y_1$ and $y_1'$),
we can build the unequal-arm Michelson observable centred on
spacecraft 1,
\begin{equation}
\fl X_1 \equiv y'_{2'} + y'_{2,2'} + y'_{3,22'} + y'_{3',322'} -
(y'_{3}  + y'_{3',3} + y'_{2',3'3} + y'_{2,2'3'3}); 
\label{eq:xobs}
\end{equation}
the analogous $X_2$ and $X_3$ are obtained by the cyclic permutations
$1 \rightarrow 2 \rightarrow 3$ and $1' \rightarrow 2' \rightarrow 3'$
(which will go without saying for all observables to follow).
Note that these $X_1$, $X_2$ and $X_3$ are usually called $X$, $Y$ and $Z$ in the TDI literature,
but here we prefer a notation that emphasizes which $y_l$ enter them.

If the missing $y_l$ are consecutive and ``codirected'' (say, $y_{2'}$
and $y_{3'}$), we can build the ``forward'' \emph{relay} observable
$U_1$ that goes through spacecraft $1$ by way of links $3$ and $2$,
\begin{equation}
\fl U_1 \equiv
y'_{1'} + y'_{1,1'} + y'_{2,11'} + y'_{3,21'1}
- (y'_{2} + y'_{3,2} + y'_{1,32} + y'_{1',132});
\end{equation}
by contrast, the ``backward'' relay observable $V_1$ goes through
spacecraft $1$ by way of $2'$ and $3'$,
\begin{equation}
\fl V_1 \equiv
y'_{1} + y'_{1',1} + y'_{3',1'1} + y'_{2',1'13'}
-(y'_{3'} + y'_{2',3'} + y'_{1',2'3'} + y'_{1,1'2'3'}).
\end{equation}

If the missing $y_l$ are consecutive, but have opposite
``directions'', we have two cases. If the available measurements are
(say) $y_1$ and $y_{1'}$, and also $y_{2'}$ and $y_{3}$ (which
``point'' towards spacecraft 1), we can build the \emph{monitor}
observable
\begin{equation}
\fl E_1 \equiv y'_{3} + y'_{1,3} + y'_{1',31} - y'_{2'} - y'_{1',2'} - y'_{1,2'1'} + y'_{2',1'1} - y'_{3,1'1};
\end{equation}
whereas if the available measurements other than $y_1$ and $y_{1'}$
are $y_2$ and $y_{3'}$ (which ``point'' away from spacecraft 1), we
can build the \emph{beacon} observable
\begin{equation}
\fl P_1 \equiv y'_{1,2} + y'_{1',21} + y'_{3',211'} - y'_{1',3'} - y'_{1,3'1'} - y'_{2,3'1'1} + y'_{2,3'} - y'_{3',2}.
\end{equation}
Again, the analogous observables $E_2$, $E_3$, $P_2$, and $P_3$ can be
obtained by cyclic permutations. The left part of table
\ref{table:failure} shows which observable is available in each case:
look at the intersection of the row and column corresponding to the
missing $y_l$.

We can now compute the noise PSDs for these observables by writing
them in the Fourier domain in terms of the $\tilde{y}_l$ and of the
complex $\Delta_l$, multiplying them by their complex conjugates, and
substituting the noise responses \eqref{eq:newynoises}.  We work in
the limit of equal armlengths (so all $\Delta_l = \exp{2 \pi \rmi f
  L}$) and assume that all noises are Gaussian and uncorrelated (i.e.,
have null cross-spectra), and have spectral densities
$S^\mathrm{op}(f)$ for the optical-path noises and $S^\mathrm{pm}(f)$
for the proof-mass noises.\footnote{For reference, the standard LISA
  model used in the Mock LISA Data Challenges \cite{mldc} has
  $(S^\mathrm{op})^{1/2} = 20 \times 10^{-12} \, \mathrm{m} \,
  \mathrm{Hz}^{-1/2}$, $(S^\mathrm{pm})^{1/2} = 3 \times 10^{-15} [1 +
  (10^{-4} \, \mathrm{Hz}/f)^2]^{1/2} \, \mathrm{m} \, \mathrm{s}^{-2}
  \, \mathrm{Hz}^{-1/2}$ and $L = 16.6782$ s.}
The resulting PSDs are
\numparts
\begin{eqnarray}
\label{eq:allspectra}
\fl S_{X_l}(f) &= 16 \, S^\mathrm{op} \sin^2 x + 16 \, S^\mathrm{pm} [3 + \cos 2 x] \sin^2 x, \label{eq:allspectrax} \\
\fl S_{U_l}(f) &= S_{V_l}(f) = 8 \, S^\mathrm{op} [4 + 4 \cos x + \cos 2 x] \sin^2 \textstyle{\frac{x}{2}} \nonumber \\ \fl & \phantom{= S_{V_l}(f) =} + 16 \, S^\mathrm{pm} [5 + 5 \cos x + 2 \cos 2 x] \sin^2 \textstyle{\frac{x}{2}}, \\
\fl S_{E_l}(f) &= S_{P_l}(f) = 8 \, S^\mathrm{op} [3 + 2 \cos x] \sin^2 \textstyle{\frac{x}{2}} + 16 \, S^\mathrm{pm} [3 + \cos x] \sin^2 \textstyle{\frac{x}{2}},
\label{eq:epspectra}
\end{eqnarray}
\endnumparts
where $x = 2 \pi f L$.
\begin{table}
\caption{Standard TDI observables available in 
various LISA failure modes. For a single lost $y_l$, all TDI 
observables on the corresponding line (or column) of the left table
are available (and any two of them are independent); for two lost
$y_l$, only the observable at the corresponding intersection is
available. For one or two lost $z_l$ on the \emph{same} spacecraft, 
the two TDI observables on the corresponding line (or column) of the 
right table is available; for two to four lost $z_l$ on \emph{two} 
spacecraft, only the observable at the intersection is available. 
For mixed failure modes, at most one $X_l$ observable is available, 
as can be seen by intersecting the two tables.\label{table:failure}}
\begin{indented}
\item[]\begin{tabular}[t]{l|l@{\hspace{4pt}}l@{\hspace{4pt}}l@{\hspace{4pt}}l@{\hspace{4pt}}l@{\hspace{4pt}}l}
\br
          & $y_1$ & $y_2$ & $y_3$ & $y_{1'}$ & $y_{2'}$ & $y_{3'}$ \\
\mr
$y_1$    &        & $V_3$  & $V_2$  & $X_1$     & $E_3$     & $P_2$     \\ 
$y_2$    & $V_3$  &        & $V_1$  & $P_3$     & $X_2$     & $E_1$     \\
$y_3$    & $V_2$  & $V_1$  &        & $E_2$     & $P_1$     & $X_3$     \\
$y_{1'}$ & $X_1$  & $P_3$  & $E_2$  &           & $U_3$     & $U_2$     \\
$y_{2'}$ & $E_3$  & $X_2$  & $P_1$  & $U_3$     &           & $U_1$     \\
$y_{3'}$ & $P_2$  & $E_1$  & $X_3$  & $U_2$     & $U_1$     &           \\
\br
\end{tabular}
\hspace{6pt}
\begin{tabular}[t]{l|l@{\hspace{4pt}}l@{\hspace{4pt}}l}
\br
                & $z_l$ on $1$ & $z_l$ on $2$ & $z_l$ on $3$ \\
\mr
$z_l$ on $1$    &              & $X_3$        & $X_2$      \\ 
$z_l$ on $2$    & $X_3$        &              & $X_1$      \\
$z_l$ on $3$    & $X_2$        & $X_1$        &            \\
\br
\end{tabular}
\end{indented}
\end{table}

\subsection{Five-link configurations}
\label{sec:fivelink}

If one $y_l$ is lost, $\mathrm{null}(D^T)$ has dimension two,
corresponding to two independent TDI observables.  Depending on which
$y_l$ is missing, a different set of the five standard observables
described in the last section can be constructed, as given by the rows
of the left part of table \ref{table:failure}.  Within each set, any
two observables can form a basis for $\mathrm{null}(D^T)$.  This means
also that any two observables can be used to re-express any other. To
find such relations, we solve linear systems such as
\begin{equation}
X_1 = c(X_1,V_2) V_2 + c(X_1,V_3) V_3,
\end{equation}
and therefore
\begin{equation}
\fl \left( \!\! \begin{array}{c}
0\\ 0\\ \Delta_{2'} - \Delta_{2'} \Delta_{3'} \Delta_{3}\\ 1 - \Delta_{3'} \Delta_{3}\\ -1 + \Delta_{2} \Delta_{2'}\\ -\Delta_{3} + \Delta_{3} \Delta_{2} \Delta_{2'}
\end{array} \!\! \right) =
\left(\!\!\begin{array}{c@{}c}
0 & 0\\
-1 + \Delta_{2'} \Delta_{2} & -\Delta_{2'} + \Delta_{2'} \Delta_{3'} \Delta_{3}\\
1 - \Delta_{2'} \Delta_{3'} \Delta_{1'} & 0\\
-\Delta_{3'} \Delta_{1'} + \Delta_{2} & -1 + \Delta_{3'} \Delta_{3}\\
0 & 1 - \Delta_{1'} \Delta_{2'} \Delta_{3'} \\
-\Delta_{1'} + \Delta_{1'} \Delta_{2'} \Delta_{2} & -\Delta_{1'} \Delta_{2'} + \Delta_{3}
\end{array}\!\!\right) \!
\left(\begin{array}{c} 
\!\!c(X_1,V_2)\!\! \\
\!\!c(X_1,V_3)\!\!
\end{array}\right),
\end{equation}
which yields
\begin{equation}
X_1 =
\frac{\Delta_{2'} - \Delta_{2'} \Delta_{3'} \Delta_{3}}{1 - \Delta_{3'} \Delta_{1'} \Delta_{2'}}
V_2 +
\frac{\Delta_{2} \Delta_{2'} - 1}{1 - \Delta_{3'} \Delta_{1'} \Delta_{2'}}
V_3
\end{equation}
and finally
\begin{equation}
X_1 - X_{1,3'1'2'} = V_{2,2'} - V_{2,2'3'3} + V_{3,22'} - V_3.
\end{equation}
Analogously,
\numparts
\begin{eqnarray}
X_{1,132} - X_1 = U_{2,3'3} - U_2 + U_{3,3} - U_{3,322'}, \\
X_{1,32} - X_{1,1'} = P_2 - P_{2,3'3} + E_{3,322'} - E_{3,3}, \\
X_{1,3'2'} - X_{1,1} = E_{2,2'3'3} - E_{2,2'} + P_3 - P_{3,22'},
\end{eqnarray}
\endnumparts
and also
\numparts
\begin{eqnarray}
V_{2,1'} - V_{2,32} = P_{2,3'1'} - P_{2,2} + E_{3,1'} - E_{3,1'2'2}, \\
V_{3,1'} - V_{3,32} = P_2 - P_{2,3'3} + E_{3,1'2'} - E_{3,3}, \\
U_{2,2'3'} - U_{2,1} = E_{2,31} - E_{2,2'} + P_3 - P_{3,22'}, \\
U_{3,2'3'} - U_{3,1} = E_{2,1} - E_{2,133'} + P_{3,21} - P_{3,3'}.
\end{eqnarray}
\endnumparts
(Since these equations cease to hold strictly if the delays become
noncommutative, here the ordering of the delay indices has little actual meaning.)

We now seek a basis of two noise-orthogonal TDI observables by
diagonalizing the cross-spectrum matrix $S_{\alpha \beta}$ of two of
the five standard observables, under the assumptions on the noises
given above.  The diagonalisation process has different (but equally
valid) results depending on which two we use, but not all choices are
equally convenient. For instance, if $y_1$ is lost and we work with
$P_2$ and $E_3$, we obtain a correlation matrix that has two equal terms
on the diagonal [see \eqref{eq:epspectra}] and \emph{complex} cross
terms
\begin{equation}
\fl S_{P_2E_3} = S^*_{E_3P_2} =
4 \, \sin \textstyle{\frac{x}{2}} \sin x \, \mathrm{e}^{\mathrm{i} x/2} [S^\mathrm{op} (3 + \mathrm{e}^{\mathrm{i} x}) + 2 \, S^\mathrm{pm} (3 - \mathrm{e}^{-\mathrm{i} x})].
\end{equation}
Unfortunately, the resulting eigenvectors are $\propto$ $E_3 -
\mathrm{e}^{\mathrm{i} \phi(f)} P_2$ and $E_3 + \mathrm{e}^{\mathrm{i}
  \phi(f)} P_2$, with $\phi(f) = \mathrm{arg}(S_{P_2E_3})$; that is,
they have \emph{complex and frequency-dependent} coefficients, whereas
we are used to orthogonal observables that are real variables and have
integer coefficients.  The same problem occurs if we diagonalize the
cross-spectrum matrix for any two out of $X_1$, $V_2$, $V_3$, $P_2$,
$E_3$. After some experimenting, we find that the linear combinations
$P_2 - X_1/2$ and $E_3 + X_1/2$ \emph{do} have a real cross-spectrum;
diagonalizing them yields the eigenvectors
\begin{equation}
A^{(1)} = (P_2 - E_3 - X_1) / \sqrt{2}, \quad
E^{(1)} = (P_2 + E_3) / \sqrt{2},
\label{eq:AE_5_link_channel_combinations}
\end{equation}
with PSDs
\numparts
\begin{eqnarray}
\fl S_{A^{(1)}} &= 8 \, S^\mathrm{op} [2 + \cos x]^2 \sin^2 \textstyle{\frac{x}{2}} + 8 \, S^\mathrm{pm} [14 + 15 \cos x + 6 \cos 2 x + \cos 3 x] \sin^2\textstyle{\frac{x}{2}}, \\
\fl S_{E^{(1)}} &= 8 \, S^\mathrm{op} [2 + \cos x]^2 \sin^2 \textstyle{\frac{x}{2}} + 32 \, S^\mathrm{pm} [2 + \cos x]^2 \sin^2\textstyle{\frac{x}{2}}.
\end{eqnarray}
\endnumparts

Working through a similar exercise when $y_{1'}$ is lost yields
\begin{equation}
A^{(1')} = (P_3 - E_2 - X_1) / \sqrt{2}, \quad
E^{(1')} = (P_3 + E_2) / \sqrt{2},
\end{equation}
with the same PSDs. Once again, cyclic index permutations will yield
orthonormal basis for the other missing $y_l$.

\subsection{Full six-link configuration}

Here $\mathrm{null}(D_6^T)$ has dimension three, and all standard
observables can be constructed.  An obvious choice is to diagonalize
the three unequal-arm Michelson observables, which have cross-spectra
\begin{equation}
\label{eq:twoxspectrum}
\fl S_{X_i X_j} =
-8 \, S^\mathrm{op} \cos x \sin^2 x - 32 \, S^\mathrm{pm} \cos x \sin^2 x \quad (i \neq j)
\end{equation}
and yield the well-known eigenvectors
\begin{eqnarray}
A^{(X)} & = (X_3 - X_1)/\sqrt{2}, \quad E^{(X)} = (X_1 - 2 X_2 + X_3)/\sqrt{6}, \nonumber \\
T^{(X)} & = (X_1 + X_2 + X_3)/\sqrt{3},
\label{eq:eigensix}
\end{eqnarray}
with PSDs
\numparts
\begin{eqnarray}
\fl S_{A^{(X)}} &= S_{E^{(X)}} = 8 \, S^\mathrm{op} [2 + \cos x] \sin^2 x + 16 \, S^\mathrm{pm} [3 + 2 \cos x + \cos 2 x] \sin^2 x, \\
\fl S_{T^{(X)}} &= 16 \, S^\mathrm{op} [1 - \cos x] \sin^2 x + 128 \,
  S^\mathrm{pm} \sin^2 x \sin^4 \textstyle{\frac{x}{2}}.
\end{eqnarray}
\endnumparts
Less well known is the fact that $E_1$, $E_2$, and $E_3$, and $P_1$, $P_2$, and $P_3$ are also
suitable bases. Since
\begin{equation}
\fl S_{P_i P_j} = S_{E_i E_j} =
4 \, S^\mathrm{op} \sin^2 x - 8 \, S^\mathrm{pm} [-1 + 2 \cos x] \sin^2 x,
\quad (i \neq j),
\end{equation}
the resulting eigenvectors have structure similar to \eqref{eq:eigensix},
\begin{eqnarray}
A^{(E)} &= (E_3 - E_1)/\sqrt{2}, \quad E^{(E)} = (E_1 - 2 E_2 + E_3)/\sqrt{6}, \nonumber \\
T^{(E)} &= (E_1 + E_2 + E_3)/\sqrt{3},
\end{eqnarray}
and
\begin{eqnarray}
A^{(P)} &= (P_3 - P_1)/\sqrt{2}, \quad E^{(P)} = (P_1 - 2 P_2 + P_3)/\sqrt{6}, \nonumber \\
T^{(P)} &= (P_1 + P_2 + P_3)/\sqrt{3},
\end{eqnarray}
with PSDs
\numparts
\begin{eqnarray}
\fl S_{A^{(E,P)}} &= S_{E^{(E,P)}} =
8 \, S^\mathrm{op} [2 + \cos x] \sin^2 \textstyle{\frac{x}{2}} + 16 \, S^\mathrm{pm} [3 + 2 \cos x + \cos 2 x] \sin^2 \textstyle{\frac{x}{2}}, \\
\fl S_{T^{(E,P)}} &=
8 \, S^\mathrm{op} [5 + 4 \cos x] \sin^2 \textstyle{\frac{x}{2}} + 32 \, S^\mathrm{pm} [5 + 4 \cos x] \sin^4 \textstyle{\frac{x}{2}}.
\end{eqnarray}
\endnumparts
By contrast, working with $U_1$, $U_2$, and $U_3$ (or $V_1$, $V_2$,
and $V_3$) leads to complex cross-spectra (although rather symmetric,
with $S_{U_1U_2} = S_{U_2U_3} = S_{U_3U_1} = S^*_{U_2U_1} =
S^*_{U_3U_2} = S^*_{U_1U_3}$), and two out of three of the resulting
eigenvectors have complex coefficients, and do not correspond to real
observables (the third eigenvectors are the again completely symmetric
$(U_1+U_2+U_3)/\sqrt{3}$ and $(V_1+V_2+V_3)/\sqrt{3}$).

The classic noise-orthogonal observables are perhaps the $A$, $E$ and
$T$ of \cite{princeetal,nayak}, written in terms of the
first-generation--TDI \emph{Sagnac} observables
\begin{equation}
\alpha = y'_{2'} + y'_{1',2'} + y'_{3',1'2'} - (y'_{3} + y'_{1,3} + y'_{2,13}),
\end{equation}
(with $\beta$ and $\gamma$ obtained by cyclical permutations) which
are a basis for $\mathrm{null}(D^T)$ only if $\Delta_l = \Delta_{l'}$,
and have PSDs
\numparts
\begin{eqnarray}
\fl S_{\alpha\alpha} = S_{\beta\beta} = S_{\gamma\gamma} = 
6 \, S^\mathrm{op} + 8 \, S^\mathrm{pm} [5 + 4 \cos x + 2 \cos 2 x] \sin^2 \textstyle{\frac{x}{2}}, \\
\fl S_{\alpha\beta} = S_{\beta\gamma} = S_{\gamma\alpha} =
2 \, S^\mathrm{op} [2 \cos x + \cos 2 x] + 4 \, S^\mathrm{pm} [\cos x - 1],
\end{eqnarray}
\endnumparts
yielding the eigenvectors
\begin{eqnarray}
A &= (\gamma - \alpha)/\sqrt{2}, \quad E = (\alpha - 2 \beta + \gamma)/\sqrt{6}, \nonumber \\
T & = (\alpha + \beta + \gamma)/\sqrt{3}
\label{eq:AET_6_link_channel_combination}
\end{eqnarray}
with PSDs
\numparts
\begin{eqnarray}
\fl S_{A} &= S_{E} =
8 \, S^\mathrm{op} [2 + \cos x] \sin^2 \textstyle{\frac{x}{2}} + 16 \, S^\mathrm{pm} [3 + 2 \cos x + \cos 2 x] \sin^2 \textstyle{\frac{x}{2}}
, \\
\fl S_{T} &=
2 \, S^\mathrm{op} [1 + 2 \cos x]^2 + 8 \, S^\mathrm{pm} \sin^2 \textstyle{\frac{3x}{2}}
.
\end{eqnarray}
\endnumparts
If $\Delta_l \neq \Delta_{l'}$ we need the more complicated
\begin{eqnarray}
\alpha_1 & = y'_{2'} + y'_{1',2'} +  y'_{3',1' 2'} +
y'_{3,3' 1' 2'} +  y'_{1,3 3' 1' 2'} +  y'_{2,1 3 3' 1' 2'} \nonumber \\
& \phantom{=}
- (y'_{3} +  y'_{1,3} +  y'_{2,1 3} +  y'_{2',2 1 3} + y'_{1',2' 2 1 3} + y'_{3',1' 2' 2 1 3})
\end{eqnarray}
(and similarly for $\alpha_2$ and $\alpha_3$), which again have \eqref{eq:eigensix}-like eigenvectors $\bar{A}$, $\bar{E}$ and $\bar{T}$, and PSDs
\begin{equation}
S_{\bar{A}} = 4 \sin^2 \textstyle{\frac{3x}{2}} \times S_{A}, \quad
S_{\bar{E}} = 4 \sin^2 \textstyle{\frac{3x}{2}} \times S_{E},
\end{equation}
since in the limit $\Delta_l \rightarrow \Delta_{l'}$ $\alpha_1 \simeq \alpha - \alpha_{123}$, and so on.

\subsection{Missing-$z_l$ configurations}

For lost $z_l$ on two spacecraft we go back to the unequal-arm
Michelson four-link scenario, as we do for certain combinations of
lost $y_l$ and $z_l$; the only scenario that remains to be covered is
one of lost $z_l$ on a single spacecraft (say, 1), where two
unequal-arm Michelson observables (in this case, $X_2$ and $X_3$) can
be constructed.  From \eqref{eq:allspectrax} and
\eqref{eq:twoxspectrum} we get the eigenvectors
\begin{equation}
A^{(z1)} = (X_2 - X_3) / \sqrt{2}, \quad  
E^{(z1)} = (X_2 + X_3) / \sqrt{2},
\end{equation}
and the PSDs
\begin{equation}
\fl S_{A^{(z1)},E^{(z1)}} = 8 \, S^\mathrm{op} [2 \pm \cos x] \sin^2 x + 8 \, S^\mathrm{pm} [6 \pm 4 \cos x + 2 \cos 2 x] \sin^2 x;
\end{equation}
in the $\pm$ signs, the $+$ refer to $A^{(z1)}$, the $-$ to $E^{(z1)}$.

\section{The LISA sensitivity with four, five and six inter-spacecraft measurements}
\label{sec:sensitivity}

To illustrate the use of noise-orthogonal TDI bases, we now compute
the sky-averaged LISA sensitivity to monochromatic, sinusoidal signals
using four, five and six $y_l$ measurements. We adopt the expressions
of Vallisneri \cite{synthlisa} for the LISA orbits and its response to
GWs, and we consider the GWs emitted by circular, nonspinning
binaries,
\begin{eqnarray}
h_+(t)      &= h (1+\cos^2\iota) \cos(2 \pi f t + \varphi_0) \nonumber \\
h_\times(t) &= -2 h \cos\iota\, \sin(2 \pi f t + \varphi_0),
\label{eq:polarizations}
\end{eqnarray}
where $\iota$ is the inclination of the binary's orbital plane with
respect to the line of sight to the Solar-system barycentre (SSB), $f$
is the frequency of the source as measured at the SSB, $\varphi_0$ is
the initial phase and $h$ is the overall GW strength.

The sensitivity to sinusoidal signals, for sources at a fixed position
in the sky, has been defined traditionally as the GW strength $h$
``required to achieve a SNR of 5 in a one-year
integration time, as a function of Fourier frequency''
\cite{whitepaper}. We follow \cite{whitepaper} in computing not the
sky-average of this position--wise sensitivity, but instead the GW
strength $h$ required to achieve an \emph{rms-averaged}
SNR of 5 for sources isotropically distributed over
the sky. However, unlike \cite{whitepaper}, where polarization states
are chosen uniformly over the abstract Poincar\'e sphere
\cite{polarization}, we consider sources isotropically distributed
over $\iota$ and over the GW polarization $\psi$ (essentially a
rotation of $h_+$ and $h_\times$ with respect to the principal axes
conventionally defined at every sky position). In fact, we find
empirically that the two distributions yield almost identical
sky-averaged sensitivity curves.

The rms sky average has the advantage that the LISA orbital motion can
be neglected, since the source distribution remains isotropic (and the
sky average invariant) as the LISA orientation changes along the year;
thus, all SNRs can be computed for a stationary LISA sitting at the
SSB. This approximation neglects the Doppler modulation of individual
sources, which changes results only marginally by smoothing out the
signals across nearby frequencies.
Following \cite{whitepaper}, we can then
operate in the frequency domain, as follows. For $10,000$ uniformly
sampled sky positions, inclinations and polarization states, we turn
$h_+(t)$ and $h_\times(t)$ into complex phasors\footnote{With
  amplitudes reduced by $1/2$ to account for a constraint of reality
  on the GWs.} that are inserted in the $y_l$ GW responses (1)--(2) of
\cite{synthlisa}, which in turn are inserted in the TDI expressions
for the observables; delays are always replaced with $\Delta_l$
factors, and evaluated in the limit of equal armlengths. The squared
modulus of the resulting phasor (which is a function of frequency
through the $\Delta_l$), multiplied by $T = 1$ year and divided by the
noise PSD, yields the $\mathrm{SNR}^2$, which can then be summed over
noise-orthogonal observables and averaged over sky position,
inclination and polarization.

\begin{figure}
\flushright
\includegraphics[width=0.80\textwidth]{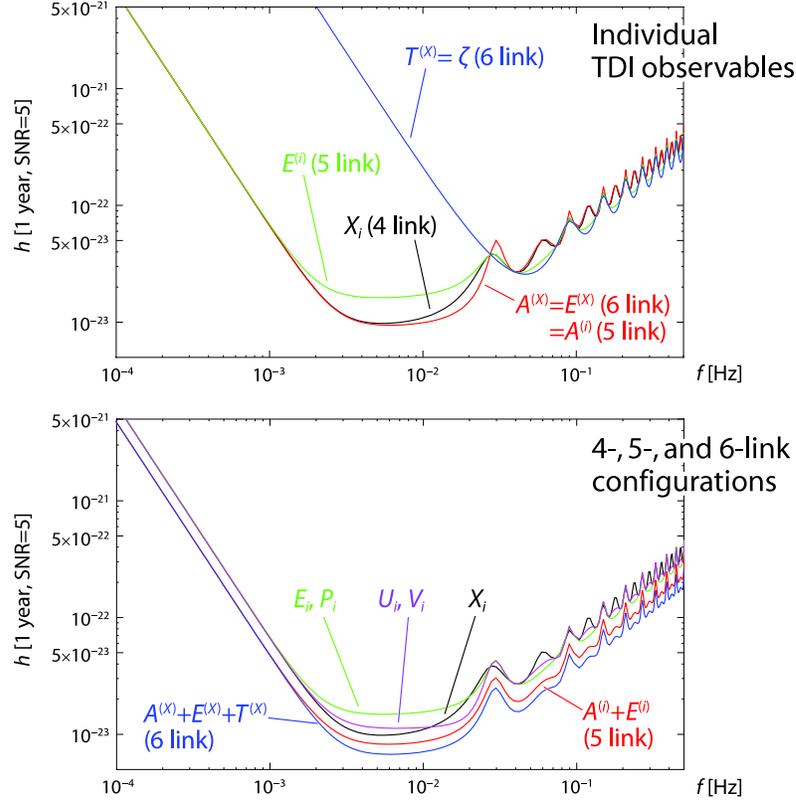}
\caption{
  Bottom panel: LISA sensitivity to monochromatic sinusoidal signals
  in the six-link, five-link, and in all four-link configurations
  (which are different for $X_i$ and for the two $E_i$/$P_i$ and
  $U_i$/$V_i$ pairs). Top panel: sensitivity of the individual
  noise-orthogonal observables in the six-link ($A^{(X)}$, $E^{(X)}$
  and $T^{(X)}$) and five-link ($A^{(i)}$, $E^{(i)}$) configurations,
  as compared to the standard unequal-arm Michelson observable $X_i$.
\label{fig:sensitivity}}
\end{figure}

Our results are shown in figure \ref{fig:sensitivity}. As plotted in
the bottom panel, the four-link configurations yield slightly
different sensitivities depending on which type of observable can be
constructed ($X_i$; $U_i$ and $V_i$, which have the same sensitivity;
or $E_i$ and $P_i$, which also have the same sensitivity); however,
they all converge in the low-frequency limit, and they essentially
agree at high frequencies.  The gain in upgrading to a five-link
configuration is $\sqrt{2}$ in the low-frequency limit, and as much as
$\sim \sqrt{3}$ at peak frequencies above 10 mHz.  The further gain in
upgrading from a four-link to a six-link configuration is again
$\sqrt{2}$ in the low-frequency limit (because the third
noise-orthogonal observable $T$ is insensitive to GWs in that limit,
as seen in the top panel), and as much as $\sim 2$ at peak frequencies
above 10 mHz.

\section{Polarization-angle estimation error with four, five and six inter-spacecraft measurements}
\label{sec:precision}

As a second example of the use of noise-orthogonal TDI bases to
characterize the LISA performance under different failure scenarios,
we now compute the expected estimation error for the GW polarization
$\psi$ of our fiducial monochromatic binaries, as predicted by the
appropriate diagonal element of the inverse Fisher matrix
(see, e.g., \cite{vallisfisher}).  One of these binaries is completely
described by 7 parameters (ecliptic latitude and longitude,
polarization, amplitude, inclination, frequency, and initial phase) so
the full Fisher matrix that we invert is $7 \times 7$. To compute its
elements, we work in the time domain with the full LISA GW response of
\cite{synthlisa} (including the amplitude and Doppler modulations due
to the LISA orbital motion), and we compute signal derivatives with
respect to source parameters by means of finite differences for very
small parameter displacements. When more than one noise-orthogonal
observable can be constructed, the corresponding
Fisher matrices are summed before inverting to yield errors.

\begin{figure}
\begin{indented}
\item[]\includegraphics[width=3.0in]{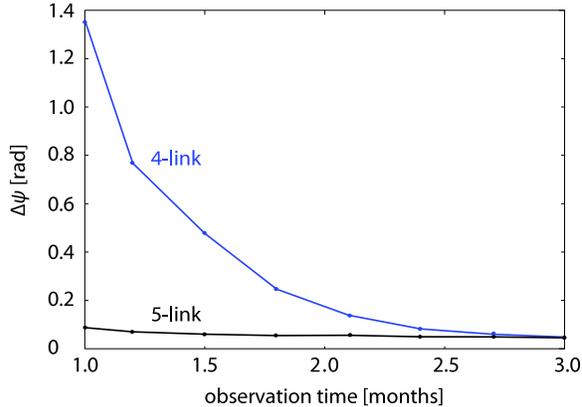}
\end{indented}
\caption{Median polarization uncertainty as a function of observation length,
  for $10,000$ sources uniformly distributed over sky
  positions, inclinations, and polarizations, as determined using
  four- and five-link LISA configurations. The signals are scaled
  separately before computing the four- and five-link errors so that
  their four- and five-link SNRs are both 10.
  \label{fig:withtime}}
\end{figure}

We consider $10,000$ binary systems spread across the LISA band, with
uniformly sampled sky positions, inclinations, and polarizations;
furthermore, we consider observations lasting one month, three months and one year,
with four-link ($X_1$), five-link ($A^{(1)} + E^{(1)}$) and six-link
($A^{(X)} + E^{(X)} + T^{(X)}$) LISA configurations. Figure \ref{fig:errors} shows the median polarization errors as a function of source frequency.
The signal amplitudes were chosen individually for each system so that it would be detected with an SNR of ten by the full six-link LISA configuration, notwithstanding the observation timespan (thus, the one-year signals are intrinsically weaker than the three-month signals, and these in turn are weaker than the one-month signals).
Several features are worth discussing:
\begin{figure}
\begin{indented}
\item[] \includegraphics[width=3.8in]{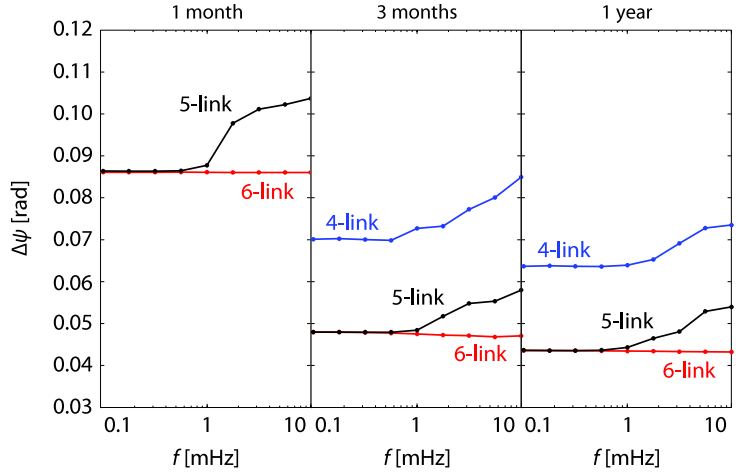}
\end{indented}
\caption{Median polarization uncertainty as a function of binary
  frequency, for $10,000$ sources uniformly distributed over sky
  positions, inclinations, and polarizations, as determined using
  four-, five- and six-link LISA configurations. The three
  panels show expected uncertainties for observations of 1 month, 3
  months and 1 year. The signals are scaled so that
  their six-link SNR is 10; thus, the four- and five-link
  uncertainties are computed for signals that would have four- and five-link SNR less than 10.
\label{fig:errors}}
\end{figure}
\begin{itemize}
\item Errors improve in the longer observations, because the LISA orbits (which have a period of one year) can then modulate the LISA response more strongly, helping to disentangle the sky-position, polarization and inclination source parameters. This improvement is most noticeable between one-month and three-month observations, and especially so for the four-link configuration, for which the one-month errors are so large ($\simeq \pi/2$) that they are off the chart in this figure, and that polarization is essentially undetermined at $\mathrm{SNR} = 10$.
\item The five- and six-link errors coincide below 1 mHz, just as the five- and six-link SNRs do. Furthermore, above 1 mHz the ratio of the six-link to five-link errors agrees (within numerical noise) with the ratio of the six-link to five-link SNRs. Recalling that Fisher-matrix expected errors scale as $\mathrm{SNR}^{-1}$, we see that the primary effect of switching from five- to six-link configurations is to improve estimation error by increasing the SNR, but not by providing complementary ``views'' of the same signal with different geometries. (Thus, if the five-link curve is renormalized by setting the five-link SNR of all binaries to 10, it collapses onto the six-link curve.)
Interestingly, the SNR-renormalized five- and six-link errors are almost constant between 0.1 and 10 mHz. This fact was already noticed by Crowder and Cornish \cite{cc2004}; unfortunately, explaining how it comes about requires delving into the analytic structure of the LISA response, and is outside the scope of this paper.
\item The four-link errors behave a bit differently: their ratio to the five- and six-link errors is also explained by the  $\mathrm{SNR}^{-1}$ scaling, \emph{but only for observations of three months or longer.} Below three months, the four-link errors get worse and worse (relative to the five- and six-link errors) as the observation time gets shorter. We observe this behaviour in figure \ref{fig:withtime}, which plots the SNR-renormalized four- and five-link errors for binaries with GW frequencies between 1 and 1.1 mHz, as functions of observation time. The curves merge at three months, but the four-link errors degrade much more quickly than the five-link errors as observations get shorter. The apparent reason is that when we reduce the observation time, the polarization and sky-position parameters become more degenerate with four links than with five; correspondingly, the parameter errors become more correlated, and therefore individually larger. Thus the additional ``view'' of the signal that is available with five links \emph{does} help the polarization error, but only for rather short observations; above three months, its effect gets washed out by the greater amount of information provided by the longer orbital baseline. 
\end{itemize}

\section{Conclusion}
\label{sec:conclusion}

We have discussed a simple framework to assess the LISA science performance (as indexed by its sensitivity to sinusoidal signals and by the inverse--Fisher-matrix parameter errors) in its full six-link configuration, and in failure modes where up to two inter-spacecraft laser measurements and up to four intra-spacecraft laser measurements become unavailable. We have described a unified, straightforward procedure to find the 
TDI observables that can be constructed in each failure mode, and to derive noise-orthonormal bases for them. In particular, we have given explicit expressions for the $A^{(i)}$ and $E^{(i)}$ orthonormal observables possible with five links, which had not previously appeared in the literature. Furthermore, we have demonstrated that 
both the SNR and the Fisher matrix are invariant under changes of TDI bases, so they depend only on which measurements are available, and can be computed for any convenient choice of observables.
As an example of the application of our framework, we have computed the LISA sensitivity to sinusoidal signals in six-link, five-link, and four-link configurations (figure \ref{fig:sensitivity}) and the expected precision for the estimation of the polarization angle of monochromatic binaries in the three configurations (figure \ref{fig:errors}). The framework assumes a ``classic'' conceptualisation of the LISA noises and measurement, but we expect that our results could be adapted easily to the emerging ``strap-down'' architecture without significant changes. The framework can also be extended trivially to derivative LISA missions that include more than three spacecraft (such as the proposed ``bow-tie'' LISA \cite{tom}) and possibly more than six inter-spacecraft measurements.

\ack

The authors are grateful to John Armstrong, Frank Estabrook, Tom Prince, Jan Harms, and to the anonymous referees for useful interactions. This work was carried out at the Jet Propulsion Laboratory, California Institute of Technology, under contract with the National Aeronautics and Space Administration. MV was supported by LISA Mission Science Office and by JPL's Human Resources Development Fund. MT was supported under research task 05-BEFS05-0014. The supercomputers used in this investigation were provided by funding from the JPL Office of the Chief Information Officer.

\appendix

\section{Multi-observable formulation of the noise inner product}
\label{app:snr}

The generalisation of the standard noise inner product to multivariate
processes (in this case, multiple TDI observables) follows from
writing the probability of the stationary Gaussian vector process
$n_\alpha$ as
\begin{equation}
\fl p(\mathbf{h}) \propto \exp \left\{ -\frac{1}{2} \int_{-\infty}^{+\infty} h_\alpha^*(f) {}^{(2)}\!S^{-1}_{\alpha \beta}(f) h_\beta(f) \, \rmd f \right\} \equiv \exp \{ -(\mathbf{h},\mathbf{h})/2 \}.
\end{equation}
which is correct since the correlations are entirely described by the cross spectrum
$\langle n_\alpha(f) n^*_\beta(f') \rangle = {}^{(2)} \! S_{\alpha \beta}(f) \delta(f - f')$.
Using the fact that $S_{\alpha \beta}$ must be Hermitian, and
furthermore that $S_{\alpha \beta}(-f) = S^*_{\alpha \beta}(f)$ (since
$n(-f) = n^*(f)$ for real noise processes), and replacing ${}^{(2)} \!
S_{\alpha \beta}(f)$ with the more familiar $S_{\alpha \beta} = 2
{}^{(2)} \! S_{\alpha \beta}$ (which becomes the one-sided spectrum
for $\alpha = \beta$), we can write
\begin{equation}
\label{eq:gensnr}
(\mathbf{g},\mathbf{h}) = 4 \, \mathrm{Re} \int_0^{+\infty} g_\alpha^*(f) S^{-1}_{\alpha \beta}(f) h_\beta(f) \, \rmd f.
\end{equation}
Note that the imaginary part of $S_{\alpha \beta}$ is crucial to
computing this inner product correctly. $S_{\alpha \beta}$ can only be
real if the correlation function $C_{\alpha \beta}(\tau) =
\int_{-\infty}^{+\infty} n_\alpha(t + \tau) n_\beta(t) \, \rmd t$ is
even, but this is never the case for TDI variables, which include
time-delayed linear combinations of different noises. Consider for
instance two noise processes $n_1(t)$ and $n_2(t) \equiv n_1(t -
\Delta t)$:
\begin{equation}
\fl C_{12}(\tau) = \! \int_{-\infty}^{+\infty} \!\! n_1(t + \tau) n_1(t - \Delta t) \, \rmd t
= \! \int_{-\infty}^{+\infty} \!\! n_1(t + \tau + \Delta t) n_1(t) \, \rmd t = C_{11}(\tau + \Delta t).
\end{equation}
Since for a stationary process $n_1$ the autocorrelation
$C_{11}(\tau)$ is even, $C_{12}(\tau)$ cannot be even. It is true
however that the imaginary part of $S_{\alpha \beta}$ may be
disregarded in computing the cross power (i.e., integrating $S_{\alpha
  \beta}$ over all frequencies) for real processes, since then all
imaginary contributions cancel between positive and negative
frequencies.

Note that in this paper we adopt the Numerical Recipes \cite{nrc}
definition of the Fourier transforms, namely $h(f) = \int h(t)
\exp^{2\pi \rmi f t} \, \rmd t$, $h(t) = \int h(f) \exp^{-2\pi \rmi f
  t} \, \rmd t$.


\section*{References}

\begin{thebibliography}{99}
%
\bibitem{PPA98} Bender P, Danzmann P and
  the LISA Study Team 1998 ``Laser Interferometer Space Antenna for the Detection of Gravitational Waves, Pre-Phase A Report'' \textbf{MPQ 233} (Garching: Max-Planck-Instit\"ut f\"ur
  Quantenoptik) 
%
\bibitem{TD05} Tinto M, and Dhurandhar S V 2005 \textit{Living Reviews
    in Relativity} \textbf{8} 4
%
\bibitem{ETA00} Estabrook F B, Tinto M and Armstrong J W 2000
  \textit{Phys. Rev.} D \textbf{62} 042002
%
\bibitem{Cutler98} Cutler C 1998 \textit{Phys. Rev.} D \textbf{57} 7089--7102
%
\bibitem{Tinto98} Tinto M 1998 \textit{Phys. Rev.} D \textbf{58}
  102001
%
\bibitem{TEA02} Tinto M, Estabrook F B and Armstrong J W 2002
  \textit{Phys. Rev.} D \textbf{65} 082003
%
\bibitem{TAE07} Tinto M, Armstrong J W and Estabrook F B 2008 \textit{Class. Quantum Grav.} \textbf{25} 015008
%
\bibitem{synthlisa} Vallisneri M 2005 \textit{Phys. Rev.} D
  \textbf{71} 022001
%
\bibitem{geotdi} Vallisneri M 2005 \textit{Phys. Rev.} D \textbf{72} 042003
%
%
\bibitem{cornhell} Cornish N J and Hellings R W 2003 \textit{Class.
Quant. Grav.} \textbf{20} 4851--4860
%
\bibitem{TEA04} Tinto M, Estabrook F B and Armstrong J W 2004
  \textit{Phys. Rev.} D \textbf{69} 082001
%
\bibitem{bonny} Schumaker B L 2007 personal communication 
%
%
%
\bibitem{golub} Golub G and van Loan C 1996 \textit{Matrix
    computations} 3rd ed. (London: Johns Hopkins Univ. Press)
%
\bibitem{tdiconnection} Romano J D and Woan G 2006 \textit{Phys. Rev.} D \textbf{73} 102001
%
\bibitem{bodewig} Bodewig E 1959 \textit{Matrix Calculus} (Amsterdam:
  North-Holland)
%
\bibitem{mldc} Arnaud K A et al 2007 \textit{Class. Quantum Grav.}
  \textbf{24} S551--S564
%
\bibitem{princeetal} Prince T A, Tinto M, Larson S L and Armstrong J W
  2002 \textit{Phys. Rev.} D \textbf{66} 122002
%
\bibitem{nayak} Nayak K R, Pai A, Dhurandhar S V and Vinet J--Y 2003 \textit{Class. Quantum Grav.} \textbf{20} 1217--1232
%
\bibitem{whitepaper} Tinto M, Estabrook F B and Armstrong J W 2002
  whitepaper on ``Time-Delay Interferometry and LISA's Sensitivity to
  Sinusoidal Gravitational Wave'' available at
  \url{www.srl.caltech.edu/lisa/mission_documents.html}
%
\bibitem{polarization} Huard S 1997 \textit{Polarization of Light}
  (Chichester, UK: Wiley)
%
\bibitem{vallisfisher} Vallisneri M 2008 \textit{Phys. Rev.} D \textit{in press} (\textit{Preprint}
  gr-qc/0703086)
%
\bibitem{cc2004} Crowder J and Cornish N J 2004 \textit{Phys. Rev.} D \textbf{70}, 082004
%
\bibitem{tom} Prince T A 2007 personal communication 
%
\bibitem{nrc} Press W H, Flannery B P, Teukolsky S A and Vetterling W
  T 1988 \textit{Numerical Recipes in C} (Cambridge: Cambridge
  University Press)
%
\end{thebibliography}
\end{document}